\documentclass[10pt,twocolumn]{article}

\usepackage{hyphenat}

\usepackage[linesnumbered,vlined]{algorithm2e}
\usepackage{booktabs} 
\usepackage{multirow}
\usepackage{multicol}
 \usepackage[table, svgnames, dvipsnames]{xcolor}
\usepackage{times}
\usepackage[pdftex]{graphicx}
\usepackage{epsfig}
\usepackage{mathtools}
\usepackage[pdftex]{geometry}
\usepackage{pdfpages}
\usepackage{indentfirst}
\usepackage[]{pdfpages}
\usepackage{booktabs}
\usepackage{multirow}
\usepackage{cite}
\usepackage[font=small,labelfont=bf]{caption}
\usepackage{hyperref}
\usepackage{dingbat}
\usepackage{bm}
\definecolor{kellygreen}{rgb}{0.5, 0.73, 0.09}
\hypersetup{
	pdfstartview={FitH},    
	pdfnewwindow=true,      
	colorlinks=true,       
	linkcolor=blue,          
	citecolor=blue,        
	filecolor=blue,      
	urlcolor=cyan           
}

\usepackage[num]{abntex2cite}
\usepackage{textcomp}

\newcommand{\balancedgrowth}{BG\xspace}
\newcommand{\dicescorecoeff}{$DSC$\xspace}

\newcommand{\growcut}{GC\xspace}
\newcommand{\hausdorffsdist}{$HD$\xspace}
\newcommand{\jaccardcoeff}{$JAC$\xspace}

\newcommand{\runtime}{$RT$\xspace}

\newcommand{\method}{3DBGrowth\xspace}

\usepackage{pgfgantt}
\usepackage{float}
\usepackage{sectsty}

\usepackage{titlesec}
\titleformat*{\section}{\large\bfseries}
\titleformat*{\subsection}{\normalsize\bfseries}
\titleformat*{\subsubsection}{\normalsize\bfseries}
\titleformat*{\paragraph}{\normalsize\bfseries}
\titleformat*{\subparagraph}{\normalsize\bfseries}
\titlespacing*{\section}
{0pt}{1ex plus 1ex minus .1ex}{1ex plus .1ex}
\titlespacing*{\subsection}
{0pt}{0.05ex plus 1ex minus .05ex}{0.05ex plus .05ex}

\usepackage{xcolor,colortbl}
\newcolumntype{g}{>{\columncolor{gray!20}}c}
\begin{document}
\twocolumn[
  \begin{@twocolumnfalse}
\begin{center}
\large{\textcolor{blue}{ 
This is a pre-print of an article published in \textbf{Computer-Based Medical Systems}. 
The final authenticated version is available online at: \url{https://doi.org/10.1109/CBMS.2019.00091}.
}}
\end{center}
\begin{center}
\LARGE{\bf \method: volumetric vertebrae segmentation and reconstruction in magnetic resonance imaging}
\end{center}
\begin{center}
	Jonathan S. Ramos$^{+}$\footnote{Corresponding author: jonathan@usp.br.}, 
Mirela T. Cazzolato$^{+}$,
Bruno S. Fai\c{c}al$^{+}$, \\
Marcello H. Nogueira-Barbosa$^{\star}$,
Caetano Traina Jr.$^{+}$ and 
Agma J. M. Traina$^{+}$\\
\vspace{0.5cm}
$^{+}$Institute of Mathematics and Computer Science (ICMC), 
University of S\~ao Paulo (USP).
\\
$^{\star}$Ribeir\~ao Preto Medical School (FMRP), 
University of S\~ao Paulo (USP). 
\end{center}

\begin{center}
\Large{\textbf{Abstract}}
\end{center}
\hrule
\vspace{0.2cm}
Segmentation of medical images is critical for making several processes of analysis and classification more reliable.
With the growing number of people presenting back pain and related problems, the semi-automatic segmentation and 3D reconstruction of vertebral bodies became even more important to support decision making.
A 3D reconstruction allows a fast and objective analysis of each vertebrae condition, which may play a major role in surgical planning and evaluation of suitable treatments.
In this paper, we propose \method, which develops a 3D reconstruction over the efficient Balanced Growth method for 2D images.
We also take advantage of the slope coefficient from the annotation time to reduce the total number of annotated slices, reducing the time spent on manual annotation.
We show experimental results on a representative dataset with 17 MRI exams demonstrating that our approach significantly outperforms the competitors and, on average, only 37\% of the total slices with vertebral body content must be annotated without losing performance/accuracy.
Compared to the state-of-the-art methods, we have achieved a Dice Score gain of over 5\% with comparable processing time.
Moreover, \method works well with imprecise seed points, which reduces the time spent on manual annotation by the specialist.
\begin{center}
\end{center}
\hrule
\vspace{0.5cm}
  \end{@twocolumnfalse}
]
\section{Introduction}
\label{sec:introduction}
Spinal diseases are increasing worldwide and can cause significant loss of function and compromise quality of life.
Surgical spinal treatments have been growing with the aging population, which requires accurate diagnosis to avoid complications~\cite{Fehlings2015}.
Many spine pathologies can be detected and diagnosed using Magnetic Resonance Imaging (MRI) exams~\cite{Rak2016, Wang2017}.
In a Computer-Aided Diagnosis (CAD) context, the segmentation of each vertebra allows a faster and more objective analysis of the vertebrae condition, aiding in the characterization and quantification of abnormalities~\cite{Hammernik2015}. 
Moreover, an accurate segmentation plays a major role and may assist the medical specialist in surgical planning and evaluation of suitable treatments~\cite{SpineWeb11}.

The manual segmentation of a vertebral body in a slice-by-slice manner may be time-consuming and prone to errors, due to inter and intra-subject variability.
\begin{figure}[!tbh]
	\centering	
\footnotesize
		\setlength{\tabcolsep}{4pt} 
	     \includegraphics[width=0.95\linewidth]{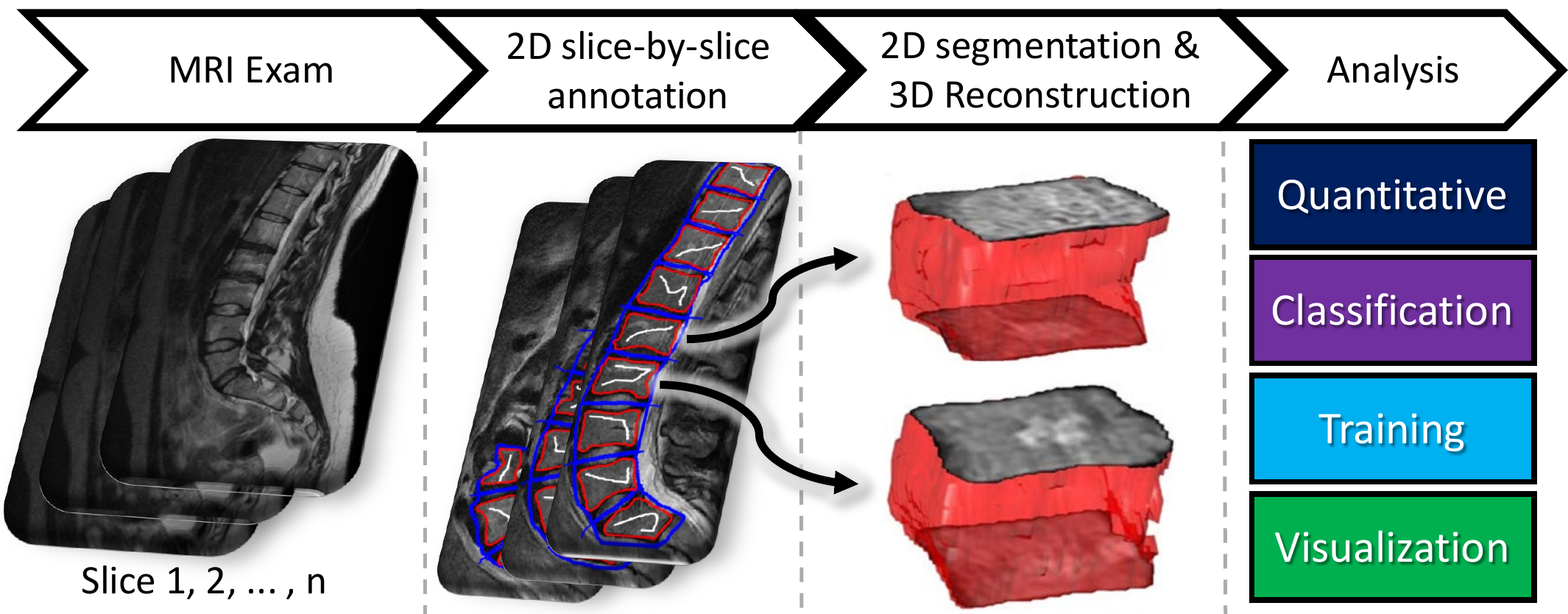} \\
	     \caption{Steps in a semi-automatic segmentation schema.}
	\label{fig:segSchema}
\end{figure}
Besides, the subjective judgment that is employed may aggregate even more inaccuracy~\cite{Gillies2016}.
Elseways, the knowledge gained over several years of expertise are incorporated.
Thus, the semi-automatic segmentation assists the specialist, leads to time savings and reduces interpretation errors~\cite{pmid29163946}.

The semi-automatic segmentation can be used in several analysis (\autoref{fig:segSchema}). \textbf{Quantitative} measures can be extracted, such as semantic and agnostic features~\cite{FERREIRAJUNIOR201823}, consequently, machine learning techniques can be applied for the \textbf{classification} of a given anomaly~\cite{casti2017, FRIGHETTOPEREIRA2016147, CazzolatoEtAl2019} or for Content-Based Image Retrieval (CBIR)~\cite{Antani2011,Antani2011-2}.
Interactive segmentation tools can be meaningful during the \textbf{training} and education of new radiologists~\cite{KarimiEtAl2018}.
Students can learn how to correctly segment each vertebra and to detect spine pathologies~\cite{StefanEtAl2018}.
%
This kind of training may avoid potential medical failures, which reduces further complications.
%
%
In general, the \textbf{visualization} of 3D human structures can be used for simulation of medical and surgical procedures~\cite{Banerjee2017}.

The GrowCut~\cite{Vezhnevets05growcut} method and its faster version, named as Fast GrowCut~\cite{FastGrowCut2014}, which presents slightly lower segmentation accuracy, have been widely used in many medical MRI exams (especially in oncology)~\cite{FERREIRAJUNIOR201823}.
The GrowCut method is based on cellular automata (analogous to a bacteria growth in biology) and works as a region-growing approach with an interactive labeling procedure~\cite{Vezhnevets05growcut}.
\begin{figure*}[!t]
	\centering	
\footnotesize
		\setlength{\tabcolsep}{4pt} 
	\includegraphics[width=0.99\linewidth]{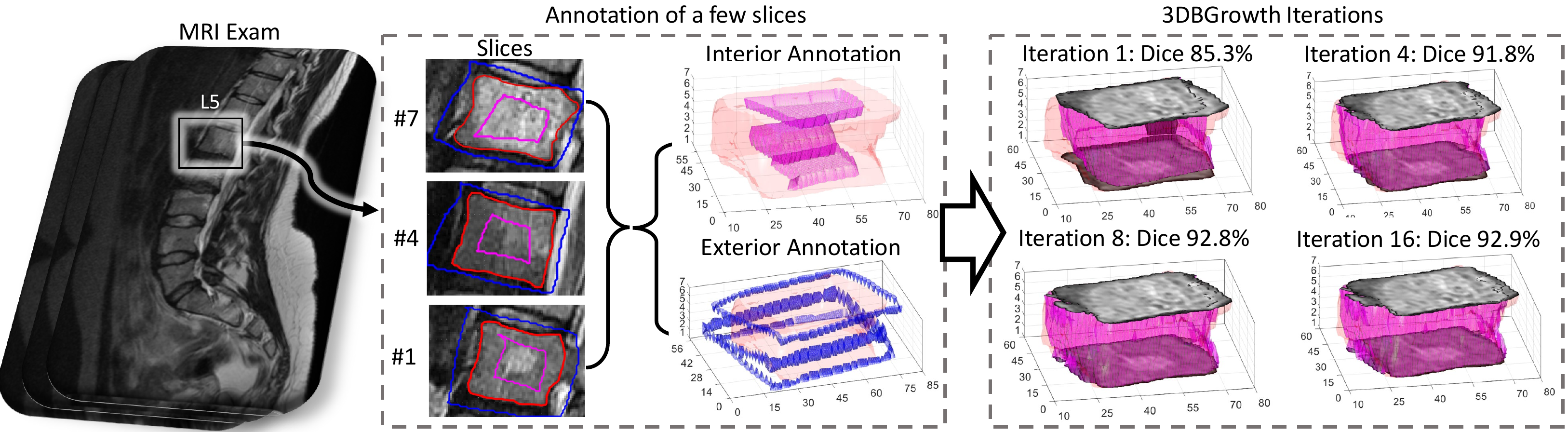}
	\caption{Examples of slices annotation for a single vertebral body (Exam AKa2, L5) and 3DBGrowth\textquotesingle s iterations. Ground-truth, interior and exterior annotation in red, magenta and blue, respectively.}
	\label{fig:an-BGrowthPipeline}
\end{figure*}

Several fully automatic vertebrae segmentation methods have been proposed~~\cite{Robert2017,Gaonkar2017}.
However, they take too much processing time, which may not suit clinical practice~\cite{HILLE201893}.
More recently, a novel approach called Balanced Growth (BGrowth)~\cite{ramos2019} has been proposed for the segmentation of crushed vertebral bodies in single slices.
Briefly, BGrowth balances the weights along the growing path of a region, so that small intensities transitions are better delineated.
%
The results achieved by BGrowth surpasses all methods from the literate, including GrowCut.
Moreover, BGrowth is able to achieve promising segmentation results even with very simple/sloppy annotation (seed points).

In this paper, we extrapolate the specialists\textquotesingle \ annotation up to a fixed limit without losing performance/accuracy, so that the total time spent on manual annotation is reduced. 
Moreover, we show how to extend BGrowth to deal with the reconstruction of volumetric exams (3D), introducing a novel method called \method.
The experimental results show that \method outperforms GrowCut, achieving an average Dice Score of 87\% while managing comparable running time. 
Moreover, the method works well even with rough seed points, which reduces the time spent on manual annotation.

The remainder of the paper is structured as follows. 
%
In \autoref{sec:proposed-method}, we present \method for the segmentation and reconstruction of vertebral bodies in volumetric MRI.
Then, in \autoref{sec:materialsMethods} we explore the materials and methods.
Next, in \autoref{sec:experiments}, we detail the experimental design, results and discussion.
Finally, \autoref{sec:conclusions} draws the conclusions.

\section{\method: The proposed method}
\label{sec:proposed-method}

The usual approach of annotating or stating seeds for segmenting medical images can be cumbersome for large 3D exams. 
Thus, this work main issue focuses on minimizing the human effort to segment and reconstruct 3D exams built on 2D slices.
As illustrated in ~\autoref{fig:an-BGrowthPipeline}, depending on the MRI exam, not all slices have to be manually annotated by the user to process the 3D reconstruction.
If the exams present a small spacing between slices (considering annotations on the sagittal plane), several slices do not need to be annotated, once they are similar.
This can be assessed by analyzing the negative slope coefficient~\cite{vanneschi2004fitness}, which gives the best trade-off between annotation time and performance measures, such as Dice Score Coefficient or Jaccard Coefficient (better explored in the next Section).

\begin{algorithm}[!thb]
\caption{\method method overview.}
\label{fig:gcPseudoCodeP}
  \normalsize
\KwIn{Image $I$ and labels matrix $L$.}
\KwOut{Segmented binary image L == 1.}
    
    $W(L \neq 0) \leftarrow 1.0$  \tcp{Initial weights}
    
\For(\tcp*[f]{For every voxel}){$\forall (i,j,z)$}{
    \For(\tcp*[f]{and its Neighbors}){$\forall (i_n, j_n, z_n)$}{
          $s \leftarrow W(i,j, z) \times \left[1 - \frac{|I(i,j, z) - I(i_n, j_n, z_n)|}{\max\limits_{\forall i,j,z}I(i,j,z)}\right]$
        
          \If{$s > W(i_n, j_n, z_n)$}{
                $W(i_n,j_n,z_n) \leftarrow (W(i_n,j_n,z_n) +s)/2$
                
                $L(i_n,j_n, z_n)\leftarrow L(i, j, z)$
        }
    }
}
\end{algorithm}

The slope between two points, $\{x_1,y_1\}$ and $\{x_2,y_2\}$, can be calculated by $\frac{\Delta y}{\Delta x} = \frac{y_2-y_1}{x_2-x_1} \ $,
which is the rate of change between the two points.
When the slope between two values of annotation time gets close to a straight horizontal line, there is no gain in annotation time, in other words, the closer the slope gets to 0, the lower the annotation time gain.
Moreover, by using a segmentation approach that does not require detailed interior/exterior annotation, such as BGrowth, the total time spent on annotation is greatly diminished.
In addition, BGrowth generally requires a simple rectangle-like annotation for the segmentation of individual vertebral bodies.
For example, only 3 out of 7 slices were annotated in~\autoref{fig:an-BGrowthPipeline}.
%
%
In average, each slice took 6.5 seconds for annotation and \method took $0.64$ seconds to process all 7 slices with only 16 iterations. 
Summing up, the whole process took $3 \times 6.5 + 0.65 = 20.65$ seconds and achieved a result close to the Ground-Truth.

In our proposed \method method (Algorithm \ref{fig:gcPseudoCodeP}), we initially consider the segmentation of foreground and background in gray-scale images.  
That is, considering a digital image $I$ and its annotations/labels as a matrix $L$, both with dimension $M \times N \times Z$, representing the number of rows, columns and slices, respectively.
Each entry in $L$ has value $-1$ (background), $0$ (unlabelled) or $1$ (foreground).

Initially, each entry in a weight matrix $W$ (with the same dimensions as $I$ and $L$) is set to $1.0$ for seeds points and $0$ otherwise (line 1).
Then, for every voxel $(i,j,z)$ and each one of its 26 neighbours $(i_n, j_n, z_n)$, a strength factor $s$ is calculated (line 4).
Here, the absolute intensity difference is normalized by the maximum intensity in the image and shifted by 1.
Finally, $s$ is multiplied by the current weight $W(i,j,z)$, which produces values within $[0,1]$.
%
If the strength $s$ is greater than the neighbour\textquotesingle s strength ($W(i_n, j_n, z_n)$, line 5), then the neighbour\textquotesingle s strength is averaged with the new strength $s$ (line 6) and its label receives the label of the voxel $(i,j,z)$ (line 7).

This process repeats until the algorithm converges or for a fixed number of iterations defined by the user.




\section{Materials and methods}
\label{sec:materialsMethods}
\begin{table}[t]
    \centering	
\footnotesize
		\setlength{\tabcolsep}{4pt} 
     \caption{Summary of measures/methods used in this work.}
    \label{tab:descriptions}
    \begin{tabular}{rl}
        \toprule
            Symbol/Acronym  & Description \\
        \midrule\midrule
               
              \dicescorecoeff & Dice Score Coefficient\\
              \jaccardcoeff & Jaccard Coefficient \\
              \hausdorffsdist & Hausdorff\textquotesingle s distance \\
              \growcut & GrowCut \\
              \balancedgrowth & 3D Balanced Growth \\
        \bottomrule
    \end{tabular}
\end{table}
The methods and measures used for comparison, as well as the computational set-up and image dataset are described as follows.

\subsection{Image Dataset}
Due to space limitations, only a meaningful dataset is presented herein, which comprises 17 anonymized MRI exams, ranging from the sacrum (S1) to the mid thoracic (T6-T12) with corresponding manual segmentations.
The exams present several health conditions, such as scoliosis, spondylolisthesis and crushed vertebra.
The exams have $3.24 \pm 0.36$ mm of slice thickness and $3.87 \pm 0.36$ mm of spacing between slices.
More information and full access to the dataset is available at~\cite{SpineWeb11}.

\subsection{Segmentation methods}
In order to evaluate the performance of \method (\balancedgrowth) in a 3D scenario, we compared it with GrowCut (\growcut), which has been widely used for the task of vertebrae segmentation~\cite{pmid29163946}. 
Since Fast GrowCut is an approximation of the original GrowCut, presenting a lower accuracy~\cite{FastGrowCut2014}, we consider only GrowCut on the experiments.
Due to the limited number of samples (exams) no deep-learning approach was applied.

\subsection{Comparison measures}
The Jaccard Coefficient (\jaccardcoeff), Dice Score Coefficient (\dicescorecoeff) and Hausdorff\textquotesingle s Distance (\hausdorffsdist) in voxels~\cite{Jaccard12similarityCoefficient,sorensen1948method} were considered.
The Jaccard (\jaccardcoeff) calculates the intersection of the manual and semi-automatic segmentation, and divides it by the union of them.
This indicates the similarity between the segmentations, in which 0 indicates no similarity and, the closer \jaccardcoeff is to 1, the more alike the segmentations~\cite{Barbieri2015}.
The Dice (\dicescorecoeff) measures the spatial overlap of several segmentations of the same object, i.e, quantifies the overlap degree between two segmented objects.
A \dicescorecoeff close to $0$ indicates very low overlap, while a \dicescorecoeff closer to 1 indicates a higher overlap.
In contrast, the Hausdorff\textquotesingle s Distance (\hausdorffsdist) indicates how far away (in voxels) the manual and semi-automatic segmentations are.
A \hausdorffsdist of 0 indicates comparable segmentations.

Table~\ref{tab:descriptions} shows a summary of the segmentation methods and comparison measures used in this work.


\begin{figure}[thb!]
	\centering	
\footnotesize
		\setlength{\tabcolsep}{4pt} 
	\begin{tabular}{ccc}
	     \includegraphics[width=0.3\linewidth]{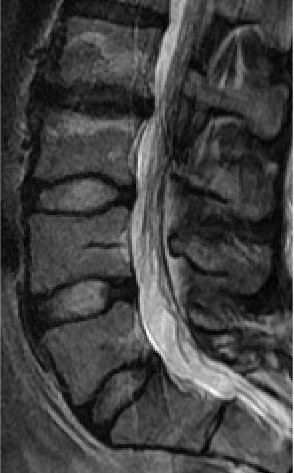} &  
	     \includegraphics[width=0.305\linewidth]{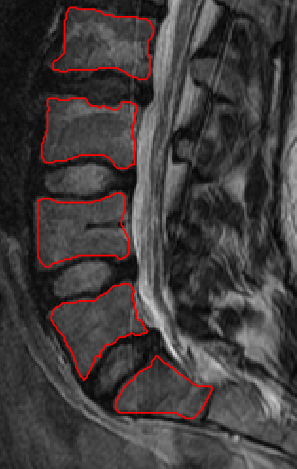} &
	     \includegraphics[width=0.3\linewidth]{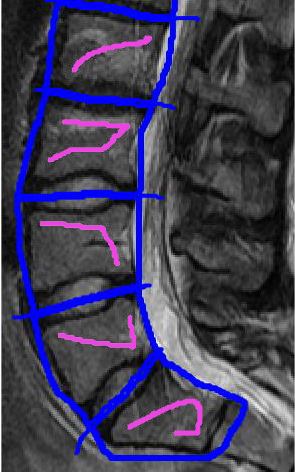} \\
	     \small{(a) Original} & \small{(b) Ground-Truth} & \small{(c) Annotation}\\
	\end{tabular}
		\caption{Example of sloppy annotation for a few vertebral bodies in one slice (Aka2, slice 8): ground-truth, interior and exterior annotations in red, magenta and blue, respectively.}
	\label{fig:an-examples}
\end{figure}

\subsection{Computational set-up}
The experiments were performed on a 2.40GHz Intel(R) Core(TM) i7 CPU and 8GB RAM machine, using Matlab(R) version 2018a.
The maximum number of iterations was set to 50 for GrowCut and \method.
No pre or post-processing technique were applied to assure the same conditions for all segmentation methods.

\section{Experiments, results and discussion}
\label{sec:experiments}

In our experimental design, we analyzed four main parts are:
(A) the performance of each segmentation method is assessed using the whole exam; 
(B) each segmentation method is tested varying the number of slices annotated for each exam; 
(C) the vertebral bodies are segmented one-by-one by each method; 
(D) a statistical test is applied to detect any significant difference between the results of the two methods.

\subsection{Exam segmentation analysis}
\label{exp:examSegmentation}

The initial interior and exterior annotation were performed in a ``sloppy'' way, i.e., no detailed boundary for accentuated curves were drawn.
In general, the annotation looks like a rectangle for the background and a simple line for the foreground (\autoref{fig:an-examples}).
For this experiment, this annotation has been performed on each slice on every exam and, to diminish computational processing, each exam is cropped using the convex hull of the exterior annotation.

Table~\ref{tb:resultsExam} shows the average Dice Score (\dicescorecoeff), Jaccard (\jaccardcoeff) and Running Time (\runtime) in seconds for each one of the 17 exams in the dataset. 
\method (\balancedgrowth) presented on average 81\% \dicescorecoeff and 68\% \jaccardcoeff while GrowCut (\growcut) presented 76\% and 61\%, respectively.
Thus, \balancedgrowth presented higher \dicescorecoeff and \jaccardcoeff percentages than \growcut for all exams, achieving up to 5\% and 7\% of \dicescorecoeff and \jaccardcoeff gain, respectively. 
Moreover, considering \dicescorecoeff and \jaccardcoeff, BG\textquotesingle s standard deviation is slightly lower.
Analyzing the Running Time (\runtime), very often, \balancedgrowth \ presented a lower average processing time than \growcut.
\begin{table}[!thb]
\centering	
\footnotesize
		\setlength{\tabcolsep}{4pt} 
\caption{Dice Score (\dicescorecoeff), Jaccard (\jaccardcoeff) and Running Time (\runtime) in seconds for \method (\balancedgrowth) and GrowCut (\growcut), considering all slices on each exam (volumetric). The best results are highlighted in bold.}
\label{tb:resultsExam}
\begin{tabular}{rccccccc} \toprule
Exam & \multicolumn{2}{c}{\dicescorecoeff (\%)} & \multicolumn{2}{c}{\jaccardcoeff (\%)} & \multicolumn{2}{c}{\runtime (s)}  \\
(\#slices) & \balancedgrowth \ \growcut & Gain & \balancedgrowth \ \growcut & Gain & \balancedgrowth & \growcut\\ \midrule\midrule
DzZ\_T1 (12) &\textbf{85}  80 & 4.86 & \textbf{74} 67 & 7.0 & \textbf{18} & 21       \\
\rowcolor{gray!10}DzZ\_T2 (12) & \textbf{82}  77 & 4.2 & \textbf{69} 63 & 5.8 & \textbf{27} & 31       \\ \
AKa2 (15)&\textbf{82}  77 & 5.17 & \textbf{69} 62 & 7.1 & 27 & 27 \\
\rowcolor{gray!10}AKa3 (15) &\textbf{78}  73 & 4.78 & \textbf{64} 58 & 6.2 & \textbf{27} & 29   \\
AKa4 (15) &\textbf{80}  73 & 7.10 & \textbf{67} 58 & 9.4 & \textbf{26} & 27    \\
\rowcolor{gray!10}AKs5 (15) &\textbf{84}  78 & 6.54 & \textbf{73} 63 & 9.2 & \textbf{23} & 24    \\
AKs6 (15) &\textbf{84}  79 & 5.44 & \textbf{73} 65 & 7.8 & \textbf{21} & 24    \\
\rowcolor{gray!10}AKs7 (15) &\textbf{80}  73 & 7.6 & \textbf{67} 57 & 9.9 & \textbf{21} & 24    \\
AKs8 (15) &\textbf{81}  78 & 3.39 & \textbf{68} 64 & 4.7 & \textbf{18} & 21    \\ 
\rowcolor{gray!10}S01 (16) &\textbf{85}  82 & 2.91 & \textbf{74} 70 & 4.3 & \textbf{44} & 50     \\ 
S02 (16) &\textbf{83}  78 & 4.97 & \textbf{70} 63 & 6.9 & \textbf{26} & 32     \\ 
\rowcolor{gray!10}F02 (18) &\textbf{78}  74 & 3.63 & \textbf{64} 59 & 4.7 & \textbf{48} & 55   \\
St1 (20) &\textbf{83}  80 & 2.73 & \textbf{71} 67 & 3.9 & \textbf{61} & 67      \\ 
\rowcolor{gray!10}F04 (23) &\textbf{78}  75 & 3.42 & \textbf{64} 60 & 4.5 & \textbf{13} & 14    \\ 
AKs3 (25) &\textbf{80}  73 & 6.42 & \textbf{66} 58 & 8.4 & 40  & \textbf{37} \\
\rowcolor{gray!10}F03 (25) &\textbf{80}  77 & 3.57 & \textbf{67} 62 & 4.8 & \textbf{08} & 09    \\
C002 (31) &\textbf{71}  65 & 5.85 & \textbf{55} 48 & 6.7 & 16 & \textbf{14} \\ \midrule\midrule
Mean &\textbf{81}  76 & 4.9 & \textbf{68} 61 & 6.6 & \textbf{27} & 30        \\
Std. dev. &\textbf{3.4}  3.9 & 1.5 & \textbf{4.7}  5.0 & 1.9 & \textbf{13.7} & 15.2  \\ \bottomrule      
\end{tabular}
\end{table}

Considering that the manual annotation of every slice in the exam is too time consuming (for this dataset, on average, 11 minutes/exam), we conducted an experiment to validate the performance of \method and GrowCut when not all slices are annotated, as explored in the next section.

\subsection{Variation on the number of annotated slices}

We used the previous experiment\textquotesingle s annotations and left a few slices without annotation: we defined a slice distance, which manages the number of non-annotated slices between two annotated slices.
For example, a slice distance of 0 implicates no slice is left without annotation.
The slice distance started at 0, increased by 1, up to 7.

\begin{figure}[!hbt]
\centering	
\footnotesize
		\setlength{\tabcolsep}{4pt} 
\begin{tabular}{c}
\includegraphics[width=0.84\linewidth]{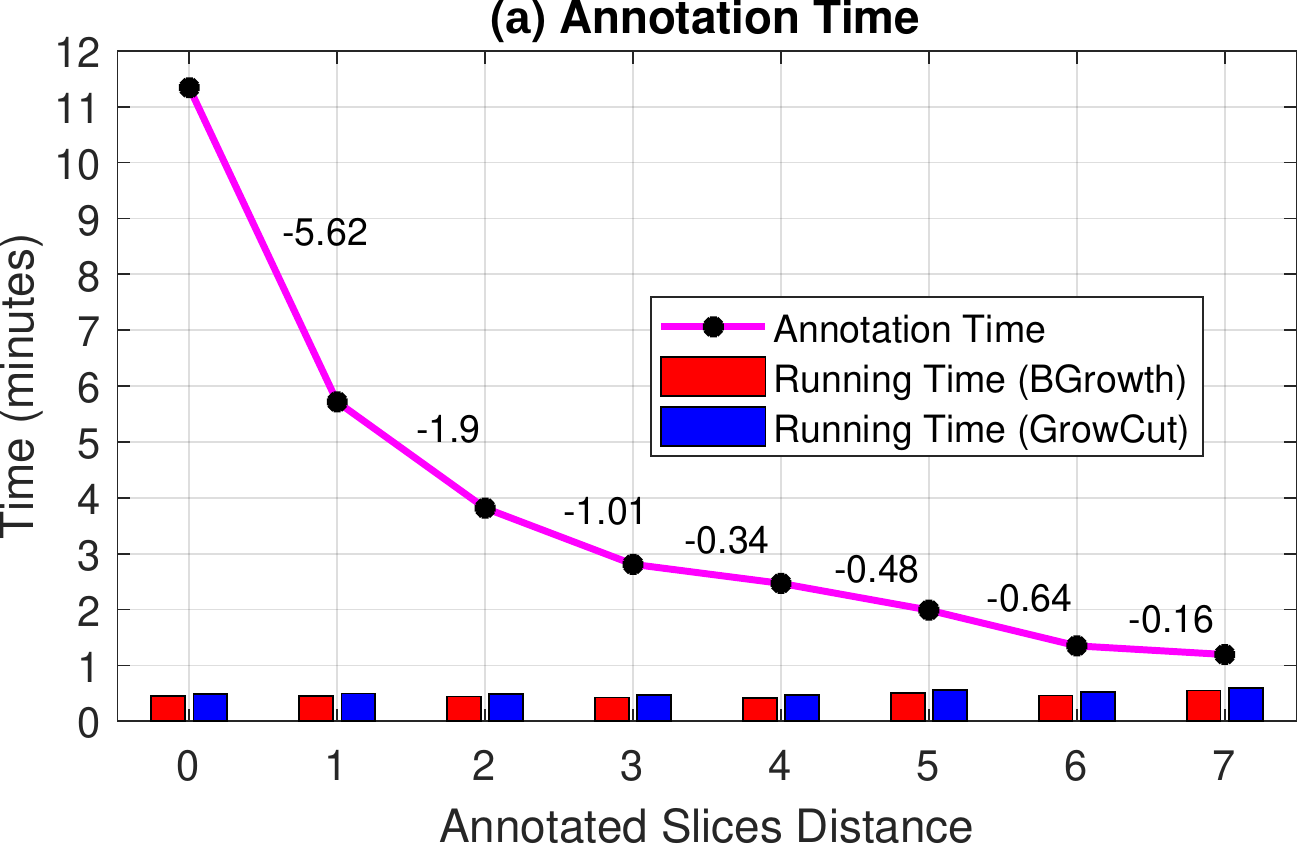}%
\\
\includegraphics[width=0.84\linewidth]{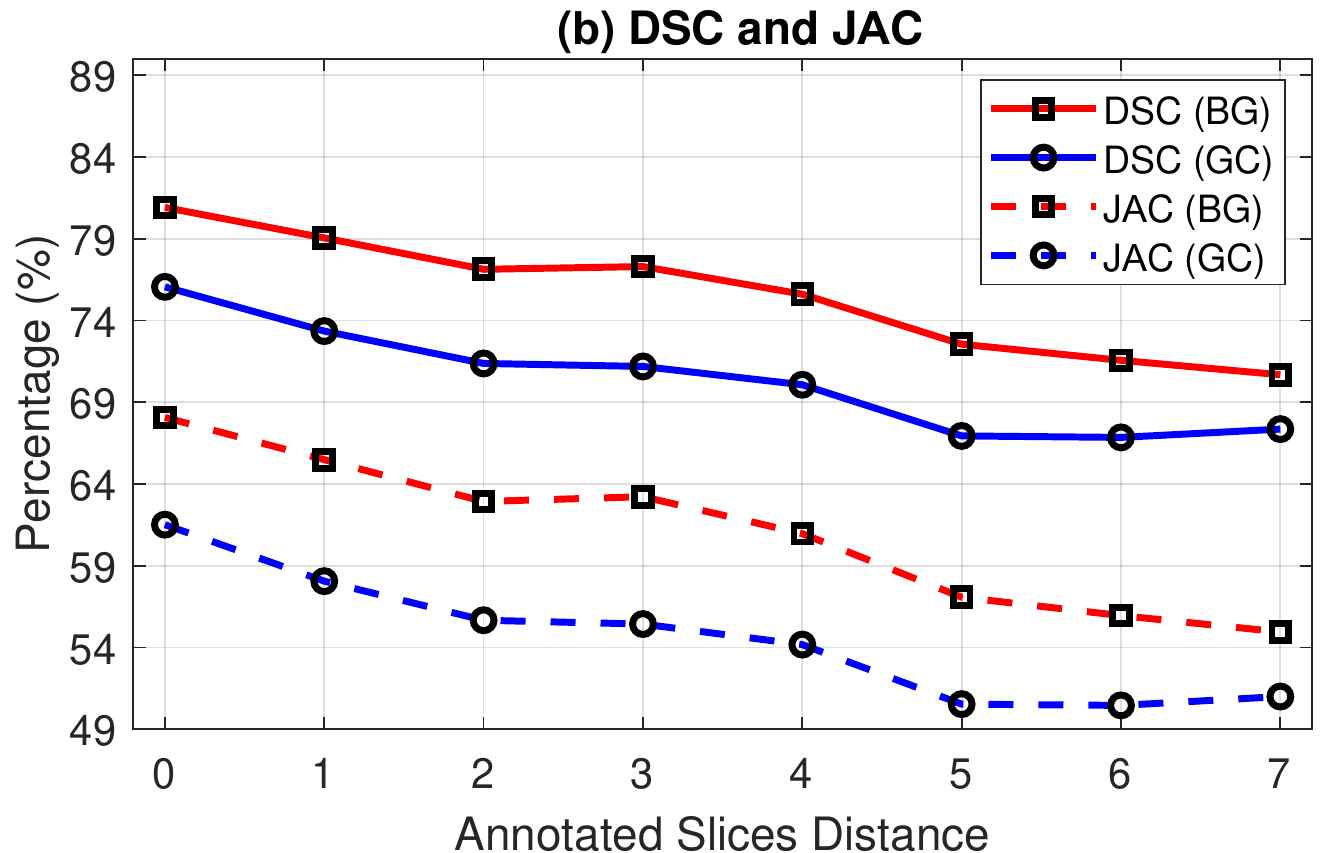}%
\end{tabular}
\caption{Quality comparison between \method and GrowCut over variations on the number of slices manually annotated: (a) annotation time and running time results; (b) Dice (\dicescorecoeff) and Jaccard (\jaccardcoeff).}
\label{fig_sim}
\end{figure}

As the slice distance increases (\autoref{fig_sim}), the average annotation time decreases and the processing time keeps almost steady for both methods.
Also, \dicescorecoeff and \jaccardcoeff drops slowly for both methods.
However, \balancedgrowth\ presented best results than \growcut for both measures.
Considering the negative slope coefficient (as discussed in Section \ref{sec:proposed-method}), highlighted over the magenta line, by using a threshold of -1, the best slice distance would be 3, which presents the best trade-off between annotation time and \dicescorecoeff/\jaccardcoeff.

\begin{figure}[!htb]
	\centering	
\footnotesize
		\setlength{\tabcolsep}{4pt} 
	\setlength\tabcolsep{1.5pt}
	\begin{tabular}{ccc}
	     \includegraphics[width=0.32\linewidth]{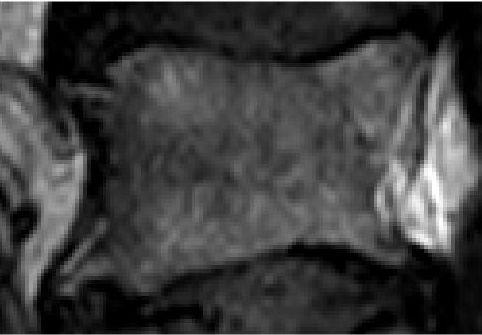}&  
	     \includegraphics[width=0.32\linewidth]{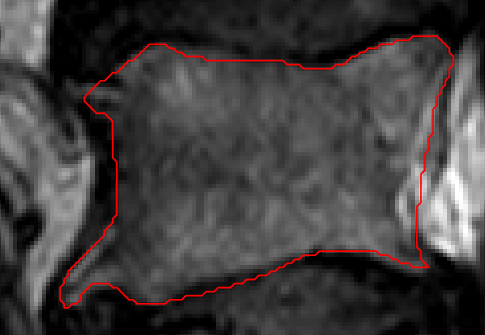} &
	     \includegraphics[width=0.32\linewidth]{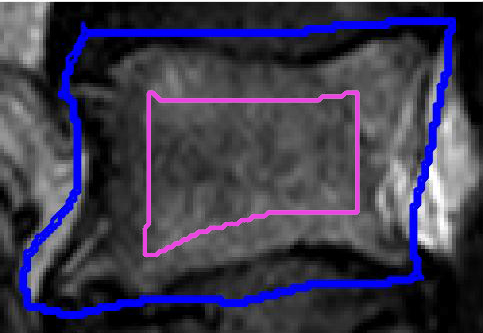} \\
	     \small{(a) Original} & \small{(b) Ground-Truth} & \small{(c) Annotation}\\
	\end{tabular}
		\caption{Example of seed points for a single vertebrae (St1, slice 10, L2): ground-truth (GT), interior and exterior annotations in red, magenta and blue, respectively.}
	\label{fig:SingleVertebraeExample}
\end{figure}

\begin{table}[thb!]
\caption{Comparison between \method (\balancedgrowth) and GrowCut (\growcut) for the Dice Score (\dicescorecoeff), Jaccard (\jaccardcoeff), Hausdorff (\hausdorffsdist) in voxels and Running Time (\runtime) in seconds. The best values are highlighted in bold.}
\label{tb:resultsVertebrae}
\centering	
\footnotesize
		\setlength{\tabcolsep}{4pt} 
\begin{tabular}{crcccc} \toprule
\multicolumn{2}{c}{} & \dicescorecoeff (\%) & \jaccardcoeff (\%) & 
\hausdorffsdist (vox.) & \runtime (s) \\

\multicolumn{2}{c}{Vertebrae}  & \balancedgrowth \growcut & \balancedgrowth \growcut & 
\balancedgrowth \growcut & \balancedgrowth \growcut\\ \midrule\midrule

& T6 & \textbf{88} \ 87 & \textbf{79} \ 78 & 
\textbf{3.16}   4.00  & \textbf{0.15} 0.16  \\ 

\rowcolor{gray!10} \cellcolor{white} & T7 & \textbf{85} \ 83 & \textbf{73} \  69 & 
\textbf{78.8}   79.1   & 7.95   \textbf{6.77} \\

& T8 & \textbf{86} \ 85 & \textbf{76} \  72 & 
 79.3   \textbf{79.2}  & \textbf{7.55}  8.37 \\

\rowcolor{gray!10} \cellcolor{white} & T9 & \textbf{81} \ 80 & 50 \ \textbf{52} & 
 80.0  \textbf{79.9}  & \textbf{7.83}  8.38 \\

 & T10& \textbf{87} \ 85 & \textbf{80}  \ 76 & 
 \textbf{26.2}   26.9 & 2.95 \textbf{2.83} \\

\rowcolor{gray!10} \cellcolor{white} & T11 & \textbf{86} \ 84 & \textbf{77} \ 73 & 
 \textbf{6.09}  6.99 & 0.91  \textbf{0.88} \\

\multirow{-7}{*}{\rotatebox{90}{Toracic}}& T12 & \textbf{89} \ 86 & \textbf{79} \ 76 & 
 \textbf{5.63}  6.88 & \textbf{1.30}  1.36 \\ \midrule

& L1 & \textbf{89} \ 87 & \textbf{78}  \ 76 & 
 \textbf{6.56}  7.31  & 1.57 \textbf{1.56} \\

\rowcolor{gray!10} \cellcolor{white} & L2 & \textbf{88} \ 86 & \textbf{79}  \ 76  & 
\textbf{6.07}   7.75 & \textbf{1.52}  1.57  \\

 & L3 & \textbf{86} \ 85 & \textbf{75} \  72 & 
\textbf{6.40}  7.49  & \textbf{1.68}  1.83 \\

\rowcolor{gray!10} \cellcolor{white}  & L4 & \textbf{88} \ 86 & \textbf{76} \  74  & 
\textbf{7.16}   7.65  & \textbf{1.77}  1.88 \\

\multirow{-5}{*}{\rotatebox{90}{Lumbar}} & L5 & \textbf{87} \ 85 & \textbf{76} \ 74 & 
\textbf{7.04}  8.42   & \textbf{2.33}  2.39  \\ \midrule

\multicolumn{2}{c}{Sacral S1} & \textbf{88} \ 86 & \textbf{79}  \ 76 &
\textbf{6.18}  7.57   & \textbf{1.74}  1.88  \\ \midrule\midrule

\multicolumn{2}{c}{Mean}    & \textbf{87} \ 85 &  \textbf{77}  \  74   & 
\textbf{7.24}  7.72 & 1.52  1.52\\ 

\multicolumn{2}{c}{Std. Dev.} &.07  \textbf{.06} & .08   .08 &
\textbf{4.85}  5.00 & 1.27  1.27 \\
\bottomrule
\end{tabular}
\end{table}

\begin{table}[thb!]
\caption{Comparison of the number of annotated slices, considering a slice distance of three.}
\label{tb:resultsVertebraeAnnotation}
\centering	
\footnotesize
		\setlength{\tabcolsep}{4pt} 
\begin{tabular}{crcccc} \toprule
\multicolumn{2}{c}{} & Slices  & out of (verte- & $ANT$\\

\multicolumn{2}{c}{Vertebrae}  & annotated & bral content) &  (seconds)\\ \midrule\midrule

& T6 & $3.0 \pm .00$	& $7.0 \pm .00$ &  $28.7 \pm .00$\\ 

\rowcolor{gray!10} \cellcolor{white} & T7 & $3.0\pm.00$	& $7.0\pm.00$ &  $32.5\pm.00$\\

& T8 & $3.0\pm.00$	& $7.0\pm.00$ & $34.6\pm11.6$\\

\rowcolor{gray!10} \cellcolor{white} & T9 & $3.0\pm.00$	& $7.0\pm.00$ & $30.0\pm6.2$\\

 & T10 & $2.5\pm.52$	& $6.7\pm3.2$ &  $25.9\pm7.1$\\

& T11 & $3.1\pm.53$	& $8.5\pm2.6$  & $30.2\pm5.8$\\

\multirow{-7}{*}{\rotatebox{90}{Toracic}}& T12 & $3.6\pm.50$	& $9.6\pm1.9$ & $34.5\pm8.1$\\ \midrule

& L1 &  $3.9\pm.78$	& $10.2\pm2.2$  & $36.8\pm9.2$\\

\rowcolor{gray!10} \cellcolor{white} & L2 &  $4.2\pm.75$	& $10.9\pm2.3$  & $38.6\pm10.1$\\

& L3 & $4.3\pm.86$	& $11.6\pm2.1$ & $40.0\pm9.5$\\

\rowcolor{gray!10} \cellcolor{white} & L4 & $4.5\pm.62$	& $12.5\pm2.8$ &  $39.3\pm6.3$\\

\multirow{-5}{*}{\rotatebox{90}{Lumbar}} & L5 & $4.5\pm.72$	& $12.5\pm3.0$ & $39.8\pm7.8$\\ \midrule

\multicolumn{2}{c}{Sacral S1} & $4.1\pm.70$	& $10.9\pm3.3$ & $35.8\pm6.3$\\\midrule\midrule

\multicolumn{2}{c}{Mean} & $4.1\pm.84$ & $10.9\pm2.9$ & $35.9\pm8.8$\\ \cmidrule{3-4}
\multicolumn{2}{c}{Annotated} & \multicolumn{2}{c}{37\%} & \\ 
\bottomrule
\end{tabular}
\end{table}
In the next Section, we conduct experiments using annotations for individual vertebral bodies.

\subsection{Individual vertebrae segmentation}
To speed-up the annotation process, we have considered a slice distance of three for this experiment.
Each vertebral body was annotated separately, as exemplified in~~\autoref{fig:SingleVertebraeExample}.
In general, both the interior and the exterior annotation looks like a rectangle and no detailed borders were drawn. 
%

%
%

\begin{figure}[bht]
	\centering	
\footnotesize
		\setlength{\tabcolsep}{4pt} 
	\setlength\tabcolsep{2pt}
	\begin{tabular}{cc}
 	\small{(a) Ground-Truth} & \small{(b) Annotation} \\
	     \includegraphics[width=0.43\linewidth]{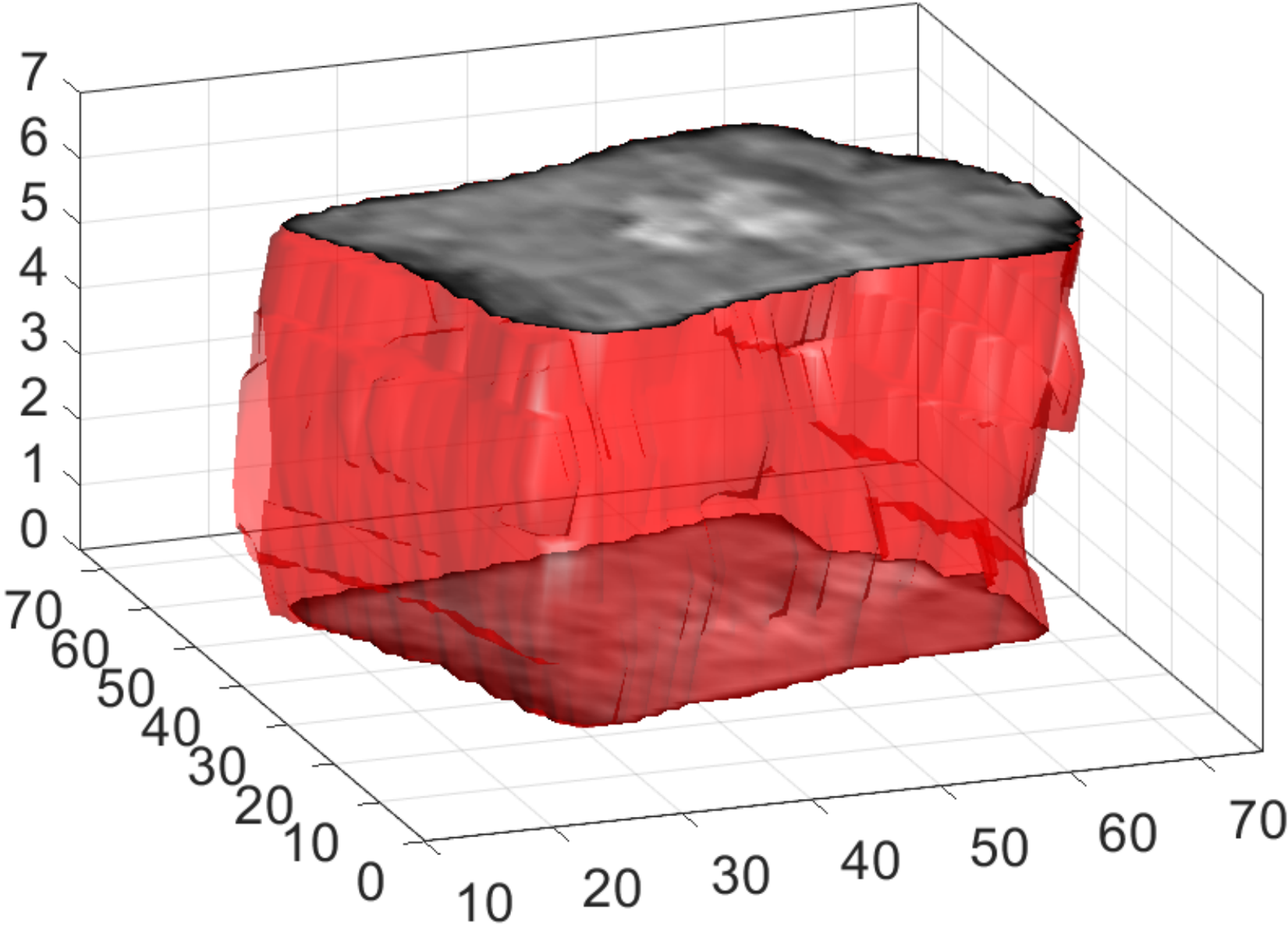} & 
	     \includegraphics[width=0.43\linewidth]{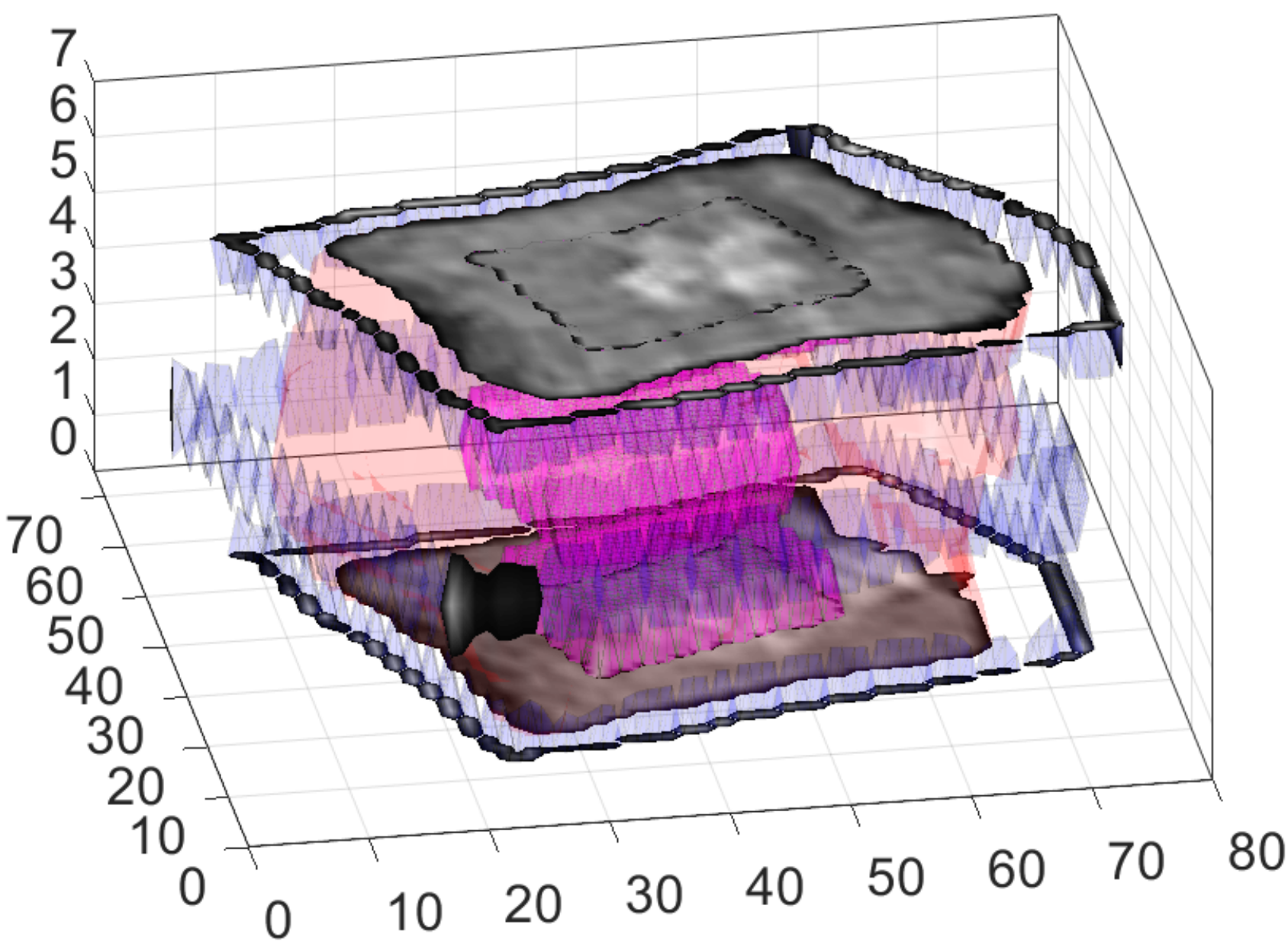} \\
    \small{(c) \method} & \small{(d) GrowCut}\\
	     \includegraphics[width=0.43\linewidth]{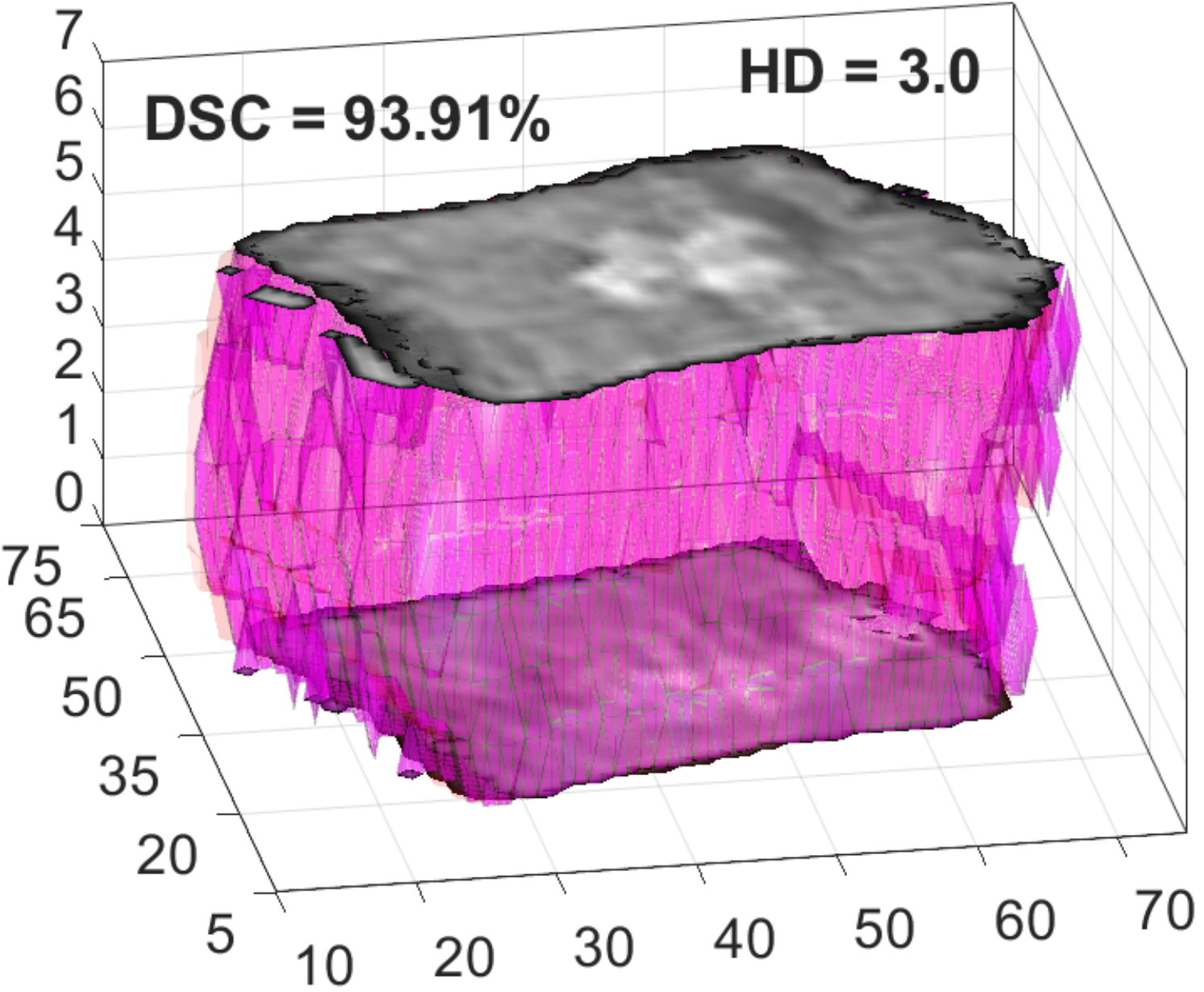} &
	     \includegraphics[width=0.43\linewidth]{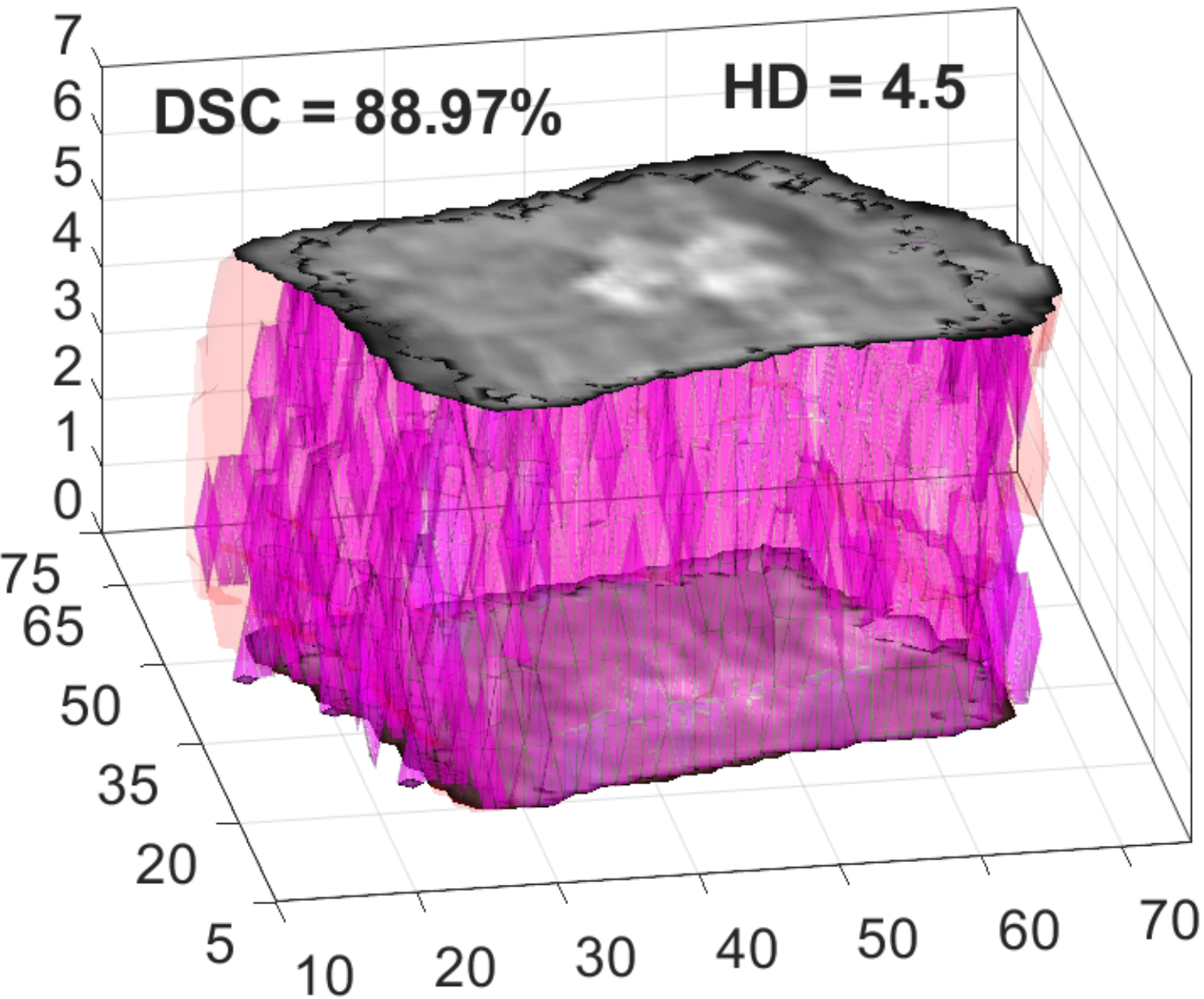} \\
	      
	\end{tabular}
		\caption{Comparison of results for L2 on exam AKa2: three slices, out of 7, were annotated.}
	\label{fig:an-3DResultsExamples}
	\vspace{-0.5cm}
\end{figure}

As reported in Table~\ref{tb:resultsVertebrae}, GC and BG presented equal mean Running Time (\runtime) and BG presented better mean DSC, JAC and HD than GrowCut. 
%
~\autoref{fig:an-3DResultsExamples} depicts the results for a single vertebral body: \balancedgrowth\ achieved the highest \dicescorecoeff and the lowest \hausdorffsdist. 
\growcut presented spiculated borders, while \balancedgrowth\ presented smooth borders (closer to the ground-truth).

Analyzing the average number of annotated slices per vertebra (Table~\ref{tb:resultsVertebraeAnnotation}), for this dataset, in average, only 37\% of the total slices with vertebral content were annotated, which speeded-up the annotation process and took, in average, 36 seconds to annotate each vertebral body.

To further investigate the results presented in Table~\ref{tb:resultsVertebrae}, we conducted a statistical test, as detailed next.

\subsection{Statistical testing}

Considering that the resulting values of each measure had several similar values, the Kolmogorov-Smirnov~\cite{kolmogorov_1951} test was applied to verify the normality of the data.
As the null hypothesis that the data follows a normal distribution was rejected for all measures, the Wilcoxon~\cite{Wilcoxon1970171} test was used to analyze if there were significant statistical differences.
In this test, the null hypothesis is that data from two dependent samples, e.g. the Dice Score (\dicescorecoeff) from \method (\balancedgrowth) and GrowCut (\growcut), were selected from populations having the same distribution, against the opposite alternative.

In the Wilcoxon test results, \method presented significantly better Dice (\dicescorecoeff), Jaccard (\jaccardcoeff) and Hausdorf\textquotesingle s Distance (\hausdorffsdist) than GrowCut.
For the Running Time (\runtime), there was no significant difference, which implicates that both methods presented comparable processing time.

\section{Conclusion}
\label{sec:conclusions}
The semi-automatic segmentation of vertebral bodies in a volumetric scenario is a challenging task, due to the large number of slices in the exams. 
To obtain a proper 3D reconstruction of the vertebrae, one has to pay attention on allowing a fast and accurate segmentation of slices.
We have investigated this challenge and used the slope coefficient of the annotation time, so that the specialists\textquotesingle\ annotations were extrapolated from a slice to its neighbours up to a given limit without losing accuracy and, at the same time, reduced the total time spent on manual annotation.

On the dataset used, on average, only 37\% of the slices with vertebral body content had to be annotated, consequently making the process faster (on average, 36 seconds for each vertebral body).
We have proposed \method method, which significantly outperforms GrowCut and keeps comparable running time. 
Moreover, \method presented the best results even with simple/sloppy seed points, which demands less effort on the annotation process.

\section*{Acknowledgment}

This study was financed in part by the Coordena\c{c}\~ao de Aperfei\c{c}oamento de Pessoal de N\'ivel Superior - Brasil (CAPES) - Finance Code 001 and grant No.: 0487/17083480, by the S\~ao Paulo Research Foundation (FAPESP, grants No. 2016/17078-0, 2017/23780-2, 2018/06228-7, 2018/24414-2), and the National Council for Scientific and Technological Development (CNPq).

\bibliographystyle{abntex2-num}
\small{\bibliography{referencias}}

\end{document}